\documentclass[aps,prd,amsmath,amssymb,onecolumn,superscriptaddress]{revtex4-2}
\usepackage{graphicx}
\usepackage{dcolumn}
\usepackage{bm}

\newcommand{\Lstar}{L_*}
\newcommand{\taustar}{\tau_*}
\newcommand{\Kmax}{K_{\mathrm{max}}}

\begin{document}

\title{Tidal Deformation Bounds and Perturbation Transfer in Bounded Curvature Spacetimes}
\author{Martin Drobczyk}
\email{martin.drobczyk@dlr.de}
\affiliation{Institute of Space Systems, German Aerospace Center (DLR), 28359 Bremen, Germany}
\date{\today}
\begin{abstract}
 We derive two model-independent results for spacetimes with globally bounded tidal fields. These are operational resolution scales of the local-inertial approximation and tidal dynamics; no spacetime discreteness is implied.
Given an invariant bound $\lambda_{\max}\le\lambda_{\rm bound}$ on the electric Riemann eigenvalues $E_{ij}\equiv R_{\hat{0}i\hat{0}j}$ along freely falling worldlines, we prove (i)~a rigorous upper bound on accumulated geodesic deviation through any bounded curvature interior, controlled by $\tau_*\equiv\lambda_{\max}^{-1/2}$, and (ii)~the existence of a critical wavenumber $k_*\sim\tau_*^{-1}$ separating adiabatic from non-adiabatic perturbation transfer through high-curvature epochs, with Bogoliubov coefficients exponentially suppressed for $k\,\tau_*\gg 1$. Both results depend only on the tidal bound (and, for mode transfer, on a mild timescale assumption for the curvature-driven effective potential) and are otherwise insensitive to metric details.
For preparation, we collect the standard operational consequences of bounded curvature, including the accuracy-dependent local-inertial domain $L_{\rm LI}(\varepsilon)\sim\sqrt{\varepsilon}\, \lambda_{\max}^{-1/2}$ and, for conformally flat cores in four dimensions, the benchmark ratio $\tau_*/L_*=24^{1/4}$ with $L_*\equiv K_{\max}^{-1/4}$. We quantify the robustness of this coefficient under departures from maximal symmetry via the Weyl-to-Kretschmann ratio $\epsilon_C$. The general framework is validated numerically in the extremal Hayward geometry.
\end{abstract}
\maketitle

\section{Introduction}
\label{sec:introduction}

Spacetime singularities in classical general relativity, where curvature invariants diverge in black hole interiors and cosmological solutions, are widely interpreted as signalling the breakdown of the classical geometric description~\cite{Penrose1965,Hawking1970}. Quantum gravity programs seek to resolve these singularities through quantization of geometry, discrete microstructure, or modified short-distance dynamics~\cite{Kiefer2007,Rovelli2004,Hossenfelder2013}. An alternative line of work assumes or dynamically enforces that spacetime curvature remains finite: Markov's limiting-density hypothesis~\cite{Markov1982}, non-singular bounce cosmologies~\cite{Mukhanov1992,Brandenberger1993}, action-level curvature constraints~\cite{Frolov2021_2D,Frolov2021_Bounce,Frolov2022_4D}, and regular black hole geometries replacing the central singularity with a high-curvature core~\cite{Bardeen1968,Hayward2006,Ansoldi2008,Frolov2016}.

Given that bounded curvature can be achieved within a classical or semiclassical framework, the present study asks: what are the precise dynamical consequences of such a bound? We formulate the hypothesis as an invariant tidal bound, $\lambda_{\max}\equiv\max_i|\lambda_i(E)|\le\lambda_{\rm bound}$, on the electric Riemann eigenvalues $E_{ij}\equiv R_{\hat{0}i\hat{0}j}$ in a parallel-transported orthonormal tetrad along freely falling worldlines. This quantity governs geodesic deviation and thus all measurable tidal accelerations~\cite{Misner1973,Wald1984}. It defines an operational proper-time scale $\tau_*\equiv\lambda_{\max}^{-1/2}$ and, through the standard Riemann normal coordinate expansion, a lower bound on the size of the local inertial frames at prescribed accuracy. A global bound on the Kretschmann scalar $K\le K_{\max}$ provides a convenient reference length $L_*\equiv K_{\max}^{-1/4}$ that becomes equivalent to a tidal bound in conformally flat, maximally symmetric cores, yielding the algebraic ratio $\tau_*/L_*=24^{1/4}$ in four dimensions. These kinematic definitions are well-known consequences of bounded curvature and are collected in section~\ref{sec:operational_scales} to fix the notation.

The central contributions of this study are two model-independent dynamical results that go beyond these definitions.

\emph{First} (section~\ref{sec:jacobi_bound}), we prove a rigorous bound on the accumulated tidal deformation along geodesics through any bounded curvature interior. If $|\lambda_i(\tau)|\le\lambda_{\max}$ along a geodesic segment of duration $\Delta\tau$, then geodesic separations satisfy
\begin{equation}
|\xi(\Delta\tau)|
\le
|\xi(0)|\,\cosh(\Delta\tau/\tau_*)
+\tau_*\,|\dot\xi(0)|\,\sinh(\Delta\tau/\tau_*),
\label{eq:jacobi_preview}
\end{equation}
with $\tau_*=\lambda_{\max}^{-1/2}$. The proof uses the Sturm comparison theorem for second-order ordinary differential equations~\cite{Hartman1964}. The bound is independent of the specific metric profile and makes precise the statement that bounded tidal eigenvalues control accumulated deformation with $\tau_*$ as the characteristic timescale. We validate this numerically using an explicit geodesic-deviation integration in the extremal Hayward geometry, where the angular stretch reaches $|\xi^{\rm ang}|/\xi_0\approx 50$ over a transit of $\Delta\tau\simeq 5.4\,\tau_*$, comfortably below the predicted ceiling of $\cosh(5.4)\approx 111$.

\emph{Second} (section~\ref{sec:perturbation_transfer}), we show that $\tau_*$ implies a critical wavenumber $k_*\sim\tau_*^{-1}$ separating adiabatic from non-adiabatic perturbation transfer through bounded curvature epochs. Modes with $k\,\tau_*\gg 1$ propagate adiabatically with exponentially suppressed Bogoliubov mixing, while modes with $k\,\tau_*\lesssim 1$ are sensitive to the detailed high-curvature profile. This result depends on the tidal bound together with a mild timescale assumption ($|\dot V|\lesssim\mathcal{O}(1)\, \lambda_{\max}^{3/2}$, automatically satisfied for single-scale curvature pulses) and is otherwise insensitive to the pulse shape, spacetime dimension, and perturbation spin. An exactly solvable P\"oschl--Teller model confirms the scaling and yields exponentially suppressed Bogoliubov coefficients with a characteristic decay scale $T_{\rm eff}\sim(2\pi\tau_*)^{-1}$.

In addition, we quantify the robustness of the benchmark coefficient $\tau_*/L_*=24^{1/4}$ under departures from maximal symmetry, parameterized by the Weyl-to-Kretschmann ratio $\epsilon_C$ (section~\ref{sec:robustness}, \ref{app:weyl}. The results apply to any framework in which curvature is bounded, including action-level limiting-curvature theories~\cite{Frolov2021_Bounce,Frolov2022_4D} and density-responsive gravity~\cite{Drobczyk2025DRG}; a quantitative calibration within specific models is deferred to future work.

The remainder of this article is organized as follows. Section~\ref{sec:related} reviews related work. Section~\ref{sec:operational_scales} presents the standard operational scales from bounded tidal fields. Section~\ref{sec:jacobi_bound} derives the model-independent deformation bound and validates it in the Hayward interior. Section~\ref{sec:perturbation_transfer} establishes the critical wavenumber for perturbation transfer. Section~\ref{sec:robustness} analyses robustness under departures from maximal symmetry. Section~\ref{sec:discussion} discusses the interpretation, limitations, and outlook.

\section{Relation to previous work}
\label{sec:related}

The hypothesis that spacetime curvature is fundamentally bounded was introduced by Markov~\cite{Markov1982} and developed into cosmological bounce models by Mukhanov and Brandenberger~\cite{Mukhanov1992,Brandenberger1993}. Recent realizations by Frolov and Zelnikov~\cite{Frolov2021_2D,Frolov2021_Bounce,Frolov2022_4D} enforce such bounds at the level of the action using auxiliary fields and Lagrange-multiplier constraints; schematically,
\begin{equation}
    S = \int d^4x \sqrt{-g} \left[ R + \sum_i \lambda_i (\chi_i - \mathcal{R}_i^2) \right],
    \label{eq:Frolov_action}
\end{equation}
where $\lambda_i$ are Lagrange multipliers, $\chi_i\geq 0$ are slack variables, and $\mathcal{R}_i$ denote curvature invariants. The precise choice of invariants and implementation differs between models. Within this broader context, the definition $L_*=K_{\max}^{-1/4}$ follows purely from dimensional analysis once a curvature bound exists and is therefore not specific to any particular framework. The present work does not propose a new action-level constraint but extracts operational consequences that follow once such a bound holds, independently of how it is generated.

The specific results derived in references~\cite{Frolov2021_Bounce,Frolov2022_4D} concern the existence and properties of cosmological bounce solutions and regular black-hole interiors within their action-level framework: demonstrating that curvature constraints can be consistently imposed, derive the modified field equations, and obtain explicit solutions.
The present work is complementary: rather than constructing solutions within a specific theory, we take the output of any such framework, a bound on the tidal eigenvalues, as input and derive two dynamical consequences (accumulated geodesic deviation and perturbation transfer) that are independent of how the bound is achieved.
The Jacobi bound (section~\ref{sec:jacobi_bound}) and the critical wavenumber $k_*$ (section~\ref{sec:perturbation_transfer}) are new results that apply equally to the Frolov--Zelnikov theory, to density-responsive gravity~\cite{Drobczyk2025DRG}, and to any other framework with bounded curvature.

Regular black hole solutions that replace the central singularity with a de~Sitter core provide the natural geometric setting for these operational statements. The Hayward metric~\cite{Hayward2006},
\begin{equation}
    f(r) = 1 - \frac{2Mr^2}{r^3 + 2M\ell^2}\,,
    \label{eq:Hayward}
\end{equation}
interpolates smoothly between a de~Sitter interior ($r\ll\ell$) and a Schwarzschild exterior ($r\gg\ell$), with the parameter $\ell$ setting the regularization scale. For maximally symmetric de~Sitter cores the Kretschmann scalar takes the value $K_{\rm dS}=24/\ell^4$, yielding the ratio $L_*/\ell=24^{-1/4}\approx 0.452$. The full curvature invariants and the proof that the Kretschmann scalar decreases monotonically from this maximum are given in~\ref{app:curvature}. This coefficient is fixed by the algebraic structure of the Riemann tensor in four dimensions and does not depend on the microscopic mechanism that produces the core. Systematic surveys of regular black hole geometries and their properties can be found in~\cite{Bardeen1968,Ansoldi2008,Frolov2016}.

\section{Operational scales from bounded tidal fields}
\label{sec:operational_scales}

This section presents the operational definitions that follow from a bound on tidal eigenvalues. The individual steps are standard consequences of Riemann normal coordinates and geodesic deviation, and we present them here to fix notation and to establish the quantities that enter the new results of sections~\ref{sec:jacobi_bound} and~\ref{sec:perturbation_transfer}.

\subsection{Tidal bound and local-inertial domain}
\label{subsec:tidal_bound}

The primary input is an invariant bound on the tidal tensor along freely falling worldlines,
\begin{equation}
\lambda_{\max}\equiv \max_i \big|\lambda_i(E_{\hat a\hat b})\big| \le \lambda_{\max}^{(\rm bound)},
\qquad
E_{\hat a\hat b}\equiv R_{\hat 0\hat a\hat 0\hat b},
\label{eq:tidal_bound}
\end{equation}
where $E_{\hat a\hat b}$ is the electric part of the Riemann tensor in a parallel-transported orthonormal tetrad. In conformally flat, maximally symmetric cores the Weyl tensor vanishes and $K=24\,\lambda_{\max}^2$, so a bound on the Kretschmann scalar $K\le K_{\max}$ is equivalent to a tidal bound. Outside this class one must bound $\lambda_{\max}$ directly.

Here and in what follows, $\lambda_{\max}$ denotes the supremum of the tidal eigenvalues along the geodesic segment under consideration, $\lambda_{\max}\equiv\sup_{\tau\in[0,\Delta\tau]} \max_i|\lambda_i(E(\tau))|$, so that $\tau_*\equiv\lambda_{\max}^{-1/2}$ is the associated segment-wise invariant timescale. When the bound is saturated (as in a de~Sitter core), the supremum coincides with the instantaneous value. The symbol $\lambda_{\max}^{(\mathrm{bound})}$ (or $\lambda_{\rm bound}$) appearing in equation~(\ref{eq:tidal_bound}) denotes an externally imposed global bound; the results below depend only on $\lambda_{\max}$ evaluated on the geodesic segment under consideration.

At any point, Riemann normal coordinates realize a local inertial frame with~\cite{Misner1973}
\begin{equation}
    g_{\mu\nu}(x) = \eta_{\mu\nu} - \frac{1}{3}R_{\mu\alpha\nu\beta}(p)\,x^\alpha x^\beta + \mathcal{O}(x^3).
    \label{eq:RNC_expansion}
\end{equation}
The local-inertial approximation holds as long as $\lambda_{\max}\,|x|^2\ll 1$. Introducing an accuracy threshold $\varepsilon$ we define the operational local-inertial radius by
\begin{equation}
\max_{\mu,\nu}\left|g_{\mu\nu}(x)-\eta_{\mu\nu}\right|\le\varepsilon
\quad \Rightarrow \quad
|x|\lesssim L_{\rm LI}(\varepsilon)\equiv \sqrt{\frac{3\varepsilon}{\lambda_{\max}}}\,.
\label{eq:LLI_def}
\end{equation}
For a maximally symmetric core, $\lambda_{\max}^{-1/2}=24^{1/4}\,L_*$ with $L_*\equiv K_{\max}^{-1/4}$, so that $L_{\rm LI}(\varepsilon)\gtrsim\sqrt{3\varepsilon}\,24^{1/4}\,L_*$. At order-unity accuracy the local-inertial domain is of size $\sim\lambda_{\max}^{-1/2}$. The estimate~(\ref{eq:LLI_def}) is an order-of-magnitude operational radius; the precise numerical prefactor depends on the chosen norm for $g_{\mu\nu}-\eta_{\mu\nu}$ and on which tensor components enter the accuracy criterion.

\begin{figure}[t]
\centering
\includegraphics[width=0.7\linewidth]{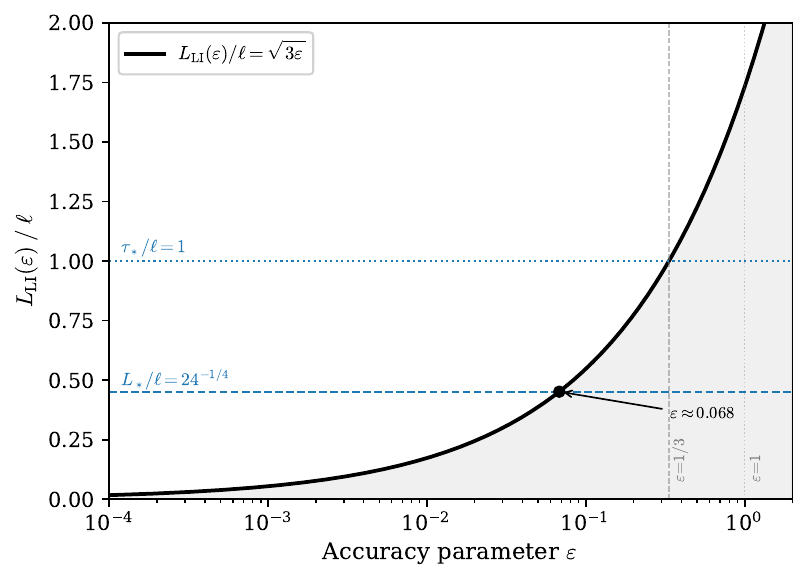}
\caption{Local-inertial domain size $L_{\rm LI}(\varepsilon)/\ell$ for a de~Sitter core with $\lambda_{\max}=\ell^{-2}$. Dashed lines mark $\varepsilon=1$ and $\varepsilon=1/3$.}
\label{fig:rnc_breakdown}
\end{figure}

Throughout this work, ``minimal'' refers to operational scales: $L_*$ is the invariant curvature length, and $\tau_*$ (defined below) is the shortest proper-time interval on which tidal evolution produces an order-unity change in geodesic separation. The underlying manifold remains smooth, proper time remains continuous, and Lorentz invariance is preserved. In quantum gravity scenarios, minimal lengths arise from modified commutators, fundamental discreteness, string-theoretic cutoffs, or noncommutative smearing~\cite{Maggiore1993,Scardigli1999,Amati1989,Rovelli2004,Bonanno2000,ReuterSaueressig2012,Nicolini2009}; the present construction is purely classical.

\subsection{Tidal timescale}
\label{subsec:tidal_timescale}

Two nearby freely falling observers separated by $\xi^\mu$ experience a relative acceleration governed by the geodesic deviation equation~\cite{Wald1984},
\begin{equation}
    \frac{D^2\xi^\mu}{d\tau^2} = -R^\mu{}_{\nu\rho\sigma}\,u^\nu\,\xi^\rho\,u^\sigma.
    \label{eq:geodesic_deviation}
\end{equation}
In the local rest space this reduces to $\ddot\xi^i=-E^i{}_j\,\xi^j$. Along an eigenmode, $\ddot\xi=-\lambda\,\xi$, yielding exponential defocusing ($\lambda<0$) or oscillatory behaviour ($\lambda>0$) with characteristic rate $|\lambda|^{1/2}$. In a de~Sitter core, $E_{ij}=-\ell^{-2}\delta_{ij}$ and $\xi(\tau)=\xi_0\,e^{\tau/\ell}$. We define the operational tidal timescale as
\begin{equation}
 \tau_* \equiv \lambda_{\max}^{-1/2}\ ,
\label{eq:taustar_boxed}
\end{equation}
the characteristic time over which geodesic separations change by order unity. The quantity $\lambda_{\max}$ is invariant under spatial rotations of the tetrad, making $\tau_*$ an invariant timescale along any freely falling worldline. For $\Delta\tau\ll\tau_*$ the accumulated distortion $\Delta\xi/\xi=\mathcal{O}(\lambda_{\max}\Delta\tau^2)\ll 1$, so tidal evolution is operationally unresolvable on that interval.

\subsection{Benchmark ratio for conformally flat cores}
\label{subsec:benchmark_ratio}

Using $L_*\equiv K_{\max}^{-1/4}$ and the maximally symmetric identity $K_{\max}=24\,\lambda_{\max}^2$ one obtains
\begin{equation}
    \frac{\tau_*}{L_*} = 24^{1/4} \approx 2.213
    \label{eq:ratio}
\end{equation}
in four dimensions (in $D$ dimensions, $\tau_*/L_*=[2D(D-1)]^{1/4}$). This coefficient is fixed by the algebraic structure of the Riemann tensor in a maximally symmetric space and receives corrections of order $\epsilon_C$ away from conformal flatness, where $\epsilon_C\equiv(C_{\mu\nu\rho\sigma}C^{\mu\nu\rho\sigma}/K)^{1/2}$ is the Weyl-to-Kretschmann ratio (section~\ref{sec:robustness}).

\section{Model-independent bound on accumulated tidal deformation}
\label{sec:jacobi_bound}

The tidal timescale $\tau_*$ introduced in section~\ref{subsec:tidal_timescale} characterizes the rate of deformation under approximately constant curvature. In a realistic interior the tidal eigenvalues vary along a geodesic, $\lambda_i(r(\tau))$, and the physically relevant question is whether the accumulated deformation of a geodesic bundle remains controlled during the full transit through the high-curvature region. We first derive a rigorous, model-independent answer and then validate it numerically in the extremal Hayward geometry.

\subsection{The deformation bound}
\label{subsec:deformation_bound}

We now derive a rigorous, model-independent upper bound on the accumulated geodesic deviation through any bounded curvature interior. The only assumption is that the tidal eigenvalues are bounded and piecewise continuous along the geodesic segment, $|\lambda(\tau)|\le\lambda_{\max}$ for $\tau\in[0,\Delta\tau]$, which ensures well-posedness of the deviation equation and the applicability of comparison arguments~\cite{Hartman1964}. For smooth regular metrics such as the Hayward geometry, $\lambda(\tau)$ is $C^\infty$ along any geodesic; therefore this regularity condition is automatically satisfied.

\paragraph{Statement.}
Let $\xi(\tau)$ satisfy the scalar geodesic deviation equation $\ddot\xi=-\lambda(\tau)\,\xi$ along a freely falling worldline, with $|\lambda(\tau)|\le\lambda_{\max}$ for all
$\tau\in[0,\Delta\tau]$. Then
\begin{equation}
 |\xi(\Delta\tau)|
  \le
  |\xi(0)|\,\cosh\!\left(\frac{\Delta\tau}{\tau_*}\right)
  +\tau_*\,|\dot\xi(0)|\,
   \sinh\!\left(\frac{\Delta\tau}{\tau_*}\right),
  \qquad
  \tau_*\equiv\lambda_{\max}^{-1/2}.
  \label{eq:jacobi_bound}
\end{equation}

\paragraph{Proof.}
Consider the comparison function $\bar\xi(\tau)=|\xi(0)|\cosh(\tau/\tau_*) +\tau_*|\dot\xi(0)|\sinh(\tau/\tau_*)$. By construction, $\ddot{\bar\xi}=\lambda_{\max}\,\bar\xi$ with $\bar\xi(0)=|\xi(0)|$ and $\dot{\bar\xi}(0)=|\dot\xi(0)|$.
Because $-\lambda(\tau)\le\lambda_{\max}$ for all $\tau$, the Sturm comparison theorem for second-order ordinary differential equations guarantees $|\xi(\tau)|\le\bar\xi(\tau)$ over the entire interval.

\paragraph{Interpretation.}
The bound depends only on the supremum $\lambda_{\max}$ of the tidal eigenvalues along the geodesic segment, and is therefore model-independent; it holds for any bounded curvature interior, regardless of the detailed profile $\lambda_i(r(\tau))$. The exponential growth rate $1/\tau_*$ is sharp in the sense that a constant de~Sitter background ($\lambda(\tau)\equiv-\lambda_{\max}$) saturates the cosh/sinh envelope.

Two consequences deserve emphasis. First, the bound specifies the dual role of $\tau_*$: it controls both the instantaneous deformation rate and the accumulated stretching over the full geodesic transit. Second, for any geometry in which the high-curvature region is traversed in a finite proper time $\Delta\tau$, the total deformation is guaranteed to remain finite, with the cosh envelope providing the worst-case ceiling.

\paragraph{Vector case and channel decoupling.}
In three spatial dimensions the full geodesic deviation equation is $\ddot{\boldsymbol{\xi}}=-E(\tau)\,\boldsymbol{\xi}$, where $E(\tau)$ is the $3\times 3$ tidal matrix. For the static, spherically symmetric geometries considered in this study, $E$ is diagonal in the parallel-transported orthonormal frame with eigenvalues $\lambda_1(\tau)$ (radial) and $\lambda_2(\tau)$ (angular, doubly degenerate). Therefore, the three tidal channels decouple exactly and the scalar bound~(\ref{eq:jacobi_bound}) applies independently to each channel with its own $\lambda_{\max}$.

In geometries where the eigenbasis rotates along the geodesic (for instance in Kerr spacetimes), inter-channel coupling prevents a direct channel-wise application. A Gronwall-type estimate for the coupled system yields
\begin{equation}
\|\boldsymbol{\xi}(\Delta\tau)\|
\;\le\;
\Big(\|\boldsymbol{\xi}(0)\|
     +\tau_*\,\|\dot{\boldsymbol{\xi}}(0)\|\Big)\,
\exp\!\Big(\Delta\tau/\tau_*\Big),
\label{eq:gronwall_fallback}
\end{equation}
obtained by rewriting the Jacobi system as a first-order system $\dot Y=B(\tau)\,Y$ with $\|B\|\le\tau_*^{-1}$ and applying Gr\"onwall's inequality. This bound is weaker than the sharp channel-wise cosh/sinh envelope but does not rely on diagonalizability of the tidal tensor and therefore applies to arbitrary geometries with bounded curvature.
 
\subsection{Validation in the extremal Hayward interior}
\label{subsec:hayward_validation}

We validate the bound by integrating the geodesic deviation equation along a radial infall geodesic through the extremal Hayward geometry ($M=3\sqrt{3}\,\ell/4$, degenerate horizon at $r_h=\sqrt{3}\,\ell$). For infall from rest at infinity ($E=1$), the radial geodesic equation reads $\dot r^2=1-f(r)$ and the principal tidal channels decouple in the parallel-transported frame,
\begin{equation}
  \ddot\xi^{\rm rad}=-\lambda_1(r)\,\xi^{\rm rad},
  \qquad
  \ddot\xi^{\rm ang}=-\lambda_2(r)\,\xi^{\rm ang},
  \label{eq:deviation_system}
\end{equation}
with $\lambda_{1,2}(r)$ given in ~(\ref{eq:tidal_eigenvalues_y}) (derived in~\ref{app:curvature}). We adopt unit initial separations with zero initial rate and integrate from $r_0=(1-10^{-3})\,r_h$ to $r_{\min}=10^{-2}\,\ell$.

\begin{figure}[t]
\centering
\includegraphics[width=\textwidth]{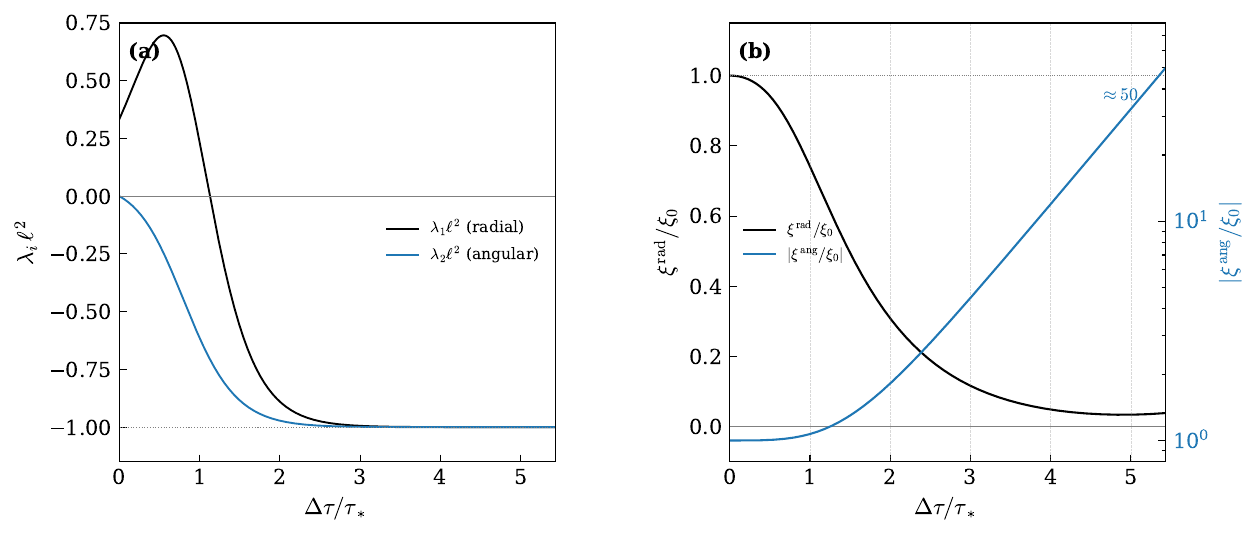}
\caption{Geodesic deviation through the extremal Hayward interior. Proper time in units of $\tau_*=\ell$. (a)~Tidal eigenvalues; grey line marks $-\ell^{-2}$. (b)~Radial separation (left axis, black): compression followed by re-expansion; angular separation (right axis, blue, log scale): monotonic growth to $|\xi^{\rm ang}|/\xi_0\approx 50$ at the cutoff.}
\label{fig:geodesic_deviation}
\end{figure}

The results are shown in Figure~\ref{fig:geodesic_deviation}. The angular component grows monotonically due to sustained defocusing ($\lambda_2<0$ throughout), while the radial component is compressed in the focusing region ($\lambda_1>0$) before transitioning to slow exponential growth in the de~Sitter core ($\lambda_1\to-\ell^{-2}$). At the integration endpoint the angular stretch is $|\xi^{\rm ang}|/\xi_0\approx 50$ and the radial separation $\xi^{\rm rad}/\xi_0\approx 0.04$. Over the first $\Delta\tau=\tau_*$ after horizon crossing, separations change by ${\sim}7$--$26\%$, consistent with $\Delta\xi/\xi\sim\mathcal{O}(\lambda_{\max}\Delta\tau^2)$. The deformation bound predicts
$|\xi|\le\cosh(5.4)\approx 111$ for the total transit
($\Delta\tau\simeq 5.4\,\tau_*$), comfortably above the
observed maximum of $50$.

\paragraph{Core asymptotics.}
In the deep core ($y\ll 1$), $\dot r^2\simeq r^2/\ell^2$ yields $r(\tau)=r_0\,e^{-\tau/\ell}$, so the center is approached only exponentially in proper time with $\Delta\tau\simeq\ell\,\ln(r_0/r_{\min})$. Because $\lambda_2\to-\ell^{-2}$ is constant in this regime, $\xi^{\rm ang}\propto e^{\tau/\ell}$ and $|\xi^{\rm ang}|/\xi_0\sim r_{\rm core}/r(\tau)$. Each additional decade of radial penetration costs $\approx 2.3\,\tau_*$ and increases the stretch by one order of magnitude.

\subsection{Geodesic completeness and contrast with Schwarzschild}
\label{subsec:completeness}

For the Hayward metric, the exponential approach $r\propto e^{-\tau/\ell}$ implies that the affine parameter extends to $\tau\to\infty$ as $r\to 0$: radial timelike geodesics are future-complete towards the core. This completeness is a consequence of the specific de~Sitter
core structure (the lapse satisfying $1-f\sim r^2/\ell^2$) and does not follow from bounded tidal eigenvalues alone; a bounded-curvature spacetime could in principle have a regular but geodesically incomplete interior if the curvature profile differs sufficiently. For the Hayward geometry, combining completeness with the deformation bound of section~\ref{subsec:deformation_bound} establishes that the geodesic deviation remains finite at every finite proper time. In contrast, in the Schwarzschild interior $|\lambda|\sim M/r^3$ diverges and the singularity is reached in a finite proper time, producing unbounded deformation. In the regular geometry the cosh bound grows exponentially with $\Delta\tau$, but because the core takes infinitely long to reach, the deformation never runs away. This geodesic completeness provides a dynamical complement to the absence of curvature singularities and the evasion of the Penrose theorem through a null energy condition violation (section~\ref{subsec:nec_epsC}).

\paragraph{Remark on generality.}
The qualitative features observed here, namely the sign change of the radial tidal eigenvalue, the monotonic angular defocusing, and the bounded total stretch, are not specific to the Hayward profile. Any regular metric interpolating between a de~Sitter core and a Schwarzschild exterior through a smooth lapse function exhibits an analogous tidal structure, because the radial eigenvalue must transition from $-\ell^{-2}$ (de~Sitter defocusing) to $+2M/r^3$ (Schwarzschild focusing) and hence must change sign in the transition region. The Bardeen metric~\cite{Bardeen1968}, for instance, shares
this feature with quantitatively similar tidal profiles in the
deep core. The deformation bound itself is of course
independent of any specific metric.

\section{Critical wavenumber for perturbation transfer}
\label{sec:perturbation_transfer}

The tidal timescale $\tau_*$ has a direct consequence for the propagation of linear perturbations through high-curvature phases. We derive a robust critical wavenumber separating adiabatic from non-adiabatic transfer and confirm the scaling with an exactly solvable model.

\subsection{Adiabaticity criterion and critical scale}
\label{subsec:adiabaticity}

Consider a mode variable $v_k(\tau)$ satisfying
\begin{equation}
  \ddot v_k + \omega_k^2(\tau)\,v_k = 0,
  \label{eq:mode_equation}
\end{equation}
where $\omega_k^2(\tau)=k^2+V(\tau)$ and $V(\tau)$ encodes the curvature driven part of the effective frequency (e.g.  $V\sim-\ddot a/a$ in a cosmological setting, or an effective potential in a black hole interior). During a bounded curvature epoch $|V|\lesssim\lambda_{\max}$ and the variation occurs on a proper-time scale $\Delta\tau\sim\tau_*$, so $|\dot V|\lesssim\lambda_{\max}^{3/2}$. The standard Wentzel--Kramers--Brillouin (WKB) adiabaticity parameter $\mathcal{A}_k\equiv|\dot\omega_k|/\omega_k^2$ scales as $(k\tau_*)^{-3}$ for $k\tau_*\gg 1$ and becomes order unity for $k\tau_*\lesssim 1$, so that modes with $k\tau_*\lesssim 1$ undergo unavoidably non-adiabatic evolution.

\paragraph{Scope of the effective potential.}
The mode equation~(\ref{eq:mode_equation}) is used here as a generic model for curvature-driven perturbation transfer, not as a literal reduction of a specific perturbation formalism. In a cosmological bounce the identification $V\sim-\ddot a/a\sim\mathcal{O}(\ell^{-2})$ is standard and $|V|\le\lambda_{\max}$ holds by construction. For black-hole perturbations the mapping is less direct: Regge--Wheeler and Zerilli potentials involve the lapse $f(r)$, its derivatives, and angular-momentum barriers; therefore, the effective $V(\tau)$ along a geodesic is a nontrivial function of the interior geometry.
In regular black-hole spacetimes with a de~Sitter core, $f(r)$, $f'(r)$, and $f''(r)$ are all bounded, and the curvature
components $\mathcal{A}$, $\mathcal{B}$, $\mathcal{F}$ (\ref{app:weyl}) satisfy $|\mathcal{A}|,\,|\mathcal{B}|,\,|\mathcal{F}| \le\mathcal{O}(\ell^{-2})$, so the bound $|V|\lesssim\lambda_{\max}$ holds parametrically. A detailed mode-by-mode verification for specific perturbation formalisms is left to future work; the present analysis establishes the scaling $k_*\sim\tau_*^{-1}$ under the stated assumptions.

\paragraph{Timescale assumption.}
To make this estimate precise we assume not only a magnitude bound on the effective potential, $|V(\tau)|\le\lambda_{\max}$, but also that its variation occurs on the natural tidal timescale,
\begin{equation}
|\dot V(\tau)|\le\alpha\,\lambda_{\max}^{3/2}
\qquad (\alpha=\mathcal{O}(1)).
\label{eq:Vdot_assumption}
\end{equation}
This is the minimal additional input needed to convert a tidal bound into an adiabaticity estimate; it is satisfied automatically when the curvature pulse has a single characteristic scale $\sim\tau_*$, in which case $\alpha$ is order unity.
Explicit evaluation gives $\alpha\approx 0.77$ for a P\"oschl--Teller profile and $\alpha\approx 0.9$--$1.8$ for the Hayward tidal channels (computed from the eigenvalue profiles of \ref{app:weyl}). Operationally,~(\ref{eq:Vdot_assumption}) states that the potential does not contain parametrically shorter features than $\tau_*$.

\paragraph{Statement.}
Under the assumptions $|V|\le\lambda_{\max}$ and $|\dot V|\le\alpha\,\lambda_{\max}^{3/2}$ with $\alpha=\mathcal{O}(1)$, the adiabaticity parameter can be bounded as follows. Because $\omega_k^2=k^2+V$ one has $\dot\omega_k=\dot V/(2\omega_k)$ and hence $\mathcal{A}_k=|\dot V|/(2\omega_k^3)$, giving
\begin{equation}
\mathcal{A}_k(\tau)
\equiv\frac{|\dot\omega_k|}{\omega_k^2}
=\frac{|\dot V|}{2\,\omega_k^3}
\lesssim
\frac{\alpha\,\lambda_{\max}^{3/2}}
     {\left(k^2+\lambda_{\max}\right)^{3/2}}\,.
\label{eq:adiabaticity_prop}
\end{equation}
Modes with $k\,\tau_*\gg 1$ therefore propagate adiabatically ($\mathcal{A}_k\ll 1$) with strongly suppressed Bogoliubov mixing, while modes with $k\,\tau_*\lesssim 1$ are generically non-adiabatic. The critical wavenumber
\begin{equation}
k_*\sim\tau_*^{-1}=\lambda_{\max}^{1/2}
\label{eq:kstar_prop}
\end{equation}
is fixed using only the tidal scale. Profile-dependent order-unity factors are encoded in $\alpha$ and in the detailed shape of $V(\tau)$.

\paragraph{Sign of the potential.}
The upper bound~(\ref{eq:adiabaticity_prop}) holds in the propagating regime $\omega_k^2>0$. For $V<0$ the effective frequency is reduced, making the adiabaticity parameter larger, but the critical scaling $k_*\sim\tau_*^{-1}$ is unchanged since $\omega_k\simeq k$ for $k\tau_*\gg 1$ regardless of the sign of $V$.

\subsection{Exact solution for a symmetric curvature pulse}
\label{subsec:poschl_teller}

To confirm the scaling quantitatively, we model a symmetric curvature pulse by a P\"oschl--Teller profile,
\begin{equation}
  V(\tau)=\frac{V_0}{\cosh^2(\tau/\tau_*)}\,,
  \qquad V_0>0,
  \label{eq:PT_potential}
\end{equation}
for which the mode equation~(\ref{eq:mode_equation}) is exactly solvable in terms of hypergeometric functions~\cite{Birrell_Davies}. Writing the standard parameterization $V_0\tau_*^2=\nu(\nu+1)$ with $\nu>0$, the exact Bogoliubov coefficient takes the closed form
\begin{equation}
|\beta_k|^2
=
\frac{\sin^2(\pi\nu)}
     {\sinh^2(\pi k\tau_*)+\sin^2(\pi\nu)}\,,
\label{eq:bogoliubov_exact}
\end{equation}
which reduces to the expected limits: $|\beta_k|^2\to\sin^2(\pi\nu)$ as $k\tau_*\to 0$ (maximal, profile-dependent mixing) and
\begin{equation}
|\beta_k|^2
\;\sim\;
4\sin^2(\pi\nu)\;e^{-2\pi k\tau_*}
\qquad (k\tau_*\gg 1),
\label{eq:beta_asymptotic}
\end{equation}
demonstrating exponential suppression above the critical scale $k_*\sim\tau_*^{-1}$. The exponential slope defines an effective temperature scale $T_{\rm eff}\sim(2\pi\tau_*)^{-1}$; ``thermal'' here refers to this high-$k$ decay rate and not to an exact Planck spectrum, which is profile- and parameter-dependent. The model-independent output is the critical scaling $k_*\sim\tau_*^{-1}$ and the exponential suppression above it, whereas profile-dependent order-unity factors reside in $\nu$.

\begin{figure}[t]
\centering
\includegraphics[width=\textwidth]{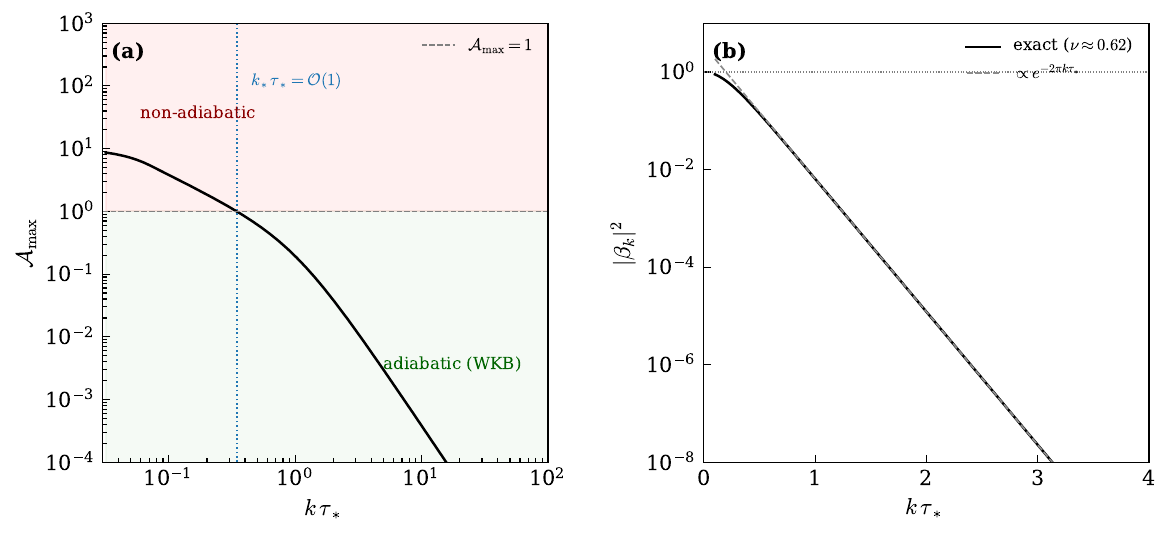}
\caption{Adiabaticity analysis for a P\"oschl--Teller curvature pulse with $V_0\tau_*^2=1$ ($\nu\approx 0.62$). (a)~Maximum WKB adiabaticity parameter $\mathcal{A}_k$; the transition at $k\tau_*=\mathcal{O}(1)$ confirms the critical scale. (b)~Exact Bogoliubov coefficient~(\ref{eq:bogoliubov_exact}) showing exponential suppression for $k\,\tau_*\gg 1$.}
\label{fig:adiabaticity}
\end{figure}

\subsection{Physical applications}
\label{subsec:applications_k}

The critical scale $k_*$ applies in two settings. In bouncing cosmologies with a bounded curvature transition, modes with $k\gg k_*$ propagate adiabatically through the high-curvature epoch and largely preserve their vacuum state, while modes with $k\lesssim k_*$ undergo significant amplification or mixing. In a de~Sitter-like bounce phase with $\lambda_{\max}\sim\ell^{-2}$ and $H_{\max}\sim\ell^{-1}$, one recovers the familiar horizon criterion $k_*/a_{\rm bounce}\sim\mathcal{O}(H_{\max})$. The same criterion governs regular-core transits: test-field modes experience non-adiabatic mixing only if $\omega_{\rm prop}\lesssim\tau_*^{-1}$.

To see this, note that in a bouncing cosmology with scale factor $a(\tau)$ and maximum Hubble rate $H_{\max}\sim\ell^{-1}$ at the bounce, the effective potential for scalar perturbations is $V\sim\ddot a/a\sim H_{\max}^2\sim\ell^{-2}\sim\lambda_{\max}$. The critical comoving wavenumber is then $k_*\sim a_{\rm bounce}\,\tau_*^{-1} \sim a_{\rm bounce}\,H_{\max}$, so the non-adiabatic band in comoving space corresponds precisely to modes whose physical wavelength exceeds the Hubble radius at the bounce, recovering the standard horizon criterion.

The scale $k_*$ provides a model-independent parametrization of which perturbation modes are sensitive to bounded curvature epochs, without fixing the resulting spectra ($n_s$, $r$, $f_{\rm NL}$), which require solving the Mukhanov-Sasaki system on a specific background. For any framework in which the core scale is of order $\ell\sim\ell_{\rm P}$, the non-adiabatic transition lies at the Planck scale.

The appearance of the scale $k_*\sim\tau_*^{-1}$ is kinematic: it follows from bounding the curvature-induced contribution to $\omega_k^2$ and its variation on the timescale $\tau_*$. The field spin and spacetime dimension affect only the detailed mapping from geometry to $V(\tau)$ and thus enter through order-unity prefactors, not through the existence or scaling of $k_*$ itself.

The connection between the deformation bound of section~\ref{sec:jacobi_bound} and the perturbation transfer criterion is worth noting: both of which are controlled by the same quantity $\tau_*$, but they address complementary aspects. The deformation bound of
section~\ref{subsec:deformation_bound} bounds the classical tidal stretching of a geodesic bundle, whereas the adiabaticity
bound of section~\ref{subsec:adiabaticity} bounds the quantum-mechanical mode mixing of field perturbations. Together, they characterize the complete operational content of a tidal timescale in bounded curvature geometries.

\section{Robustness of the operational scales}
\label{sec:robustness}

The coefficient $\tau_*/L_*=24^{1/4}$ is exact only when the curvature bound is saturated in a locally maximally symmetric, conformally flat core. In this section we quantify how the operational interpretation persists when this idealization is relaxed.

The algebraic decomposition of the Riemann tensor in four dimensions reads
\begin{equation}
 R_{\mu\nu\rho\sigma}
=
C_{\mu\nu\rho\sigma}
+\frac{1}{2}\left(g_{\mu\rho}S_{\nu\sigma}-g_{\mu\sigma}S_{\nu\rho}-g_{\nu\rho}S_{\mu\sigma}+g_{\nu\sigma}S_{\mu\rho}\right)
-\frac{R}{12}\left(g_{\mu\rho}g_{\nu\sigma}-g_{\mu\sigma}g_{\nu\rho}\right),
\label{eq:riemann_decomp}
\end{equation}
where $C_{\mu\nu\rho\sigma}$ is the Weyl tensor and $S_{\mu\nu}\equiv R_{\mu\nu}-\frac{1}{4}Rg_{\mu\nu}$ is the trace-free Ricci tensor. The Kretschmann scalar decomposes as $K=C_{\mu\nu\rho\sigma}C^{\mu\nu\rho\sigma}+2R_{\mu\nu}R^{\mu\nu}-\frac{1}{3}R^2$. In a maximally symmetric core $C_{\mu\nu\rho\sigma}=0$ and $S_{\mu\nu}=0$, forcing a unique relation between $K$ and the tidal eigenvalues. Away from maximal symmetry, $K$ alone does not determine the tidal spectrum, but the operational construction remains well-defined when phrased in terms of $\lambda_{\max}$ as in equation~(\ref{eq:taustar_boxed}).

We parameterize departures from conformal flatness by the dimensionless ratio
\begin{equation}
\epsilon_C \equiv \left(\frac{C_{\mu\nu\rho\sigma}C^{\mu\nu\rho\sigma}}{K}\right)^{1/2},
\qquad 0\le \epsilon_C \le 1,
\label{eq:epsC_def}
\end{equation}
which vanishes identically for conformally flat cores. For $\epsilon_C\ll 1$ the tidal eigenvalues can be viewed as a degenerate maximally symmetric spectrum plus a perturbation controlled by the Weyl curvature, yielding
\begin{align}
\lambda_{\max}
&= \lambda_{\max}^{(0)}\left[1+\mathcal{O}(\epsilon_C)\right],
\\
\frac{\tau_*}{L_*}
&= 24^{1/4}\left[1+\mathcal{O}(\epsilon_C)\right]
\quad
\text{(near-conformally-flat core)}.
\label{eq:ratio_perturb}
\end{align}
The notation $\mathcal{O}(\epsilon_C)$ indicates a parametric correlation rather than a convergent perturbative expansion: the
tidal spectrum deviates from the maximally symmetric benchmark by an amount that scales with the Weyl fraction. For $\epsilon_C\ll 1$ (deep core) the deviation is linear and small ($<2\%$ for $\epsilon_C<0.07$); at intermediate radii where $\epsilon_C\sim 0.2$--$0.3$ the deviation reaches ${\sim}26\%$ (Table~\ref{tab:robustness}), reflecting a non-perturbative departure from conformal flatness. The benchmark $24^{1/4}$ is therefore reliable in the deep interior where the bounded-curvature hypothesis is most relevant, but should not be applied without correction at intermediate radii.

An explicit evaluation of $\epsilon_C(r)$ for the Hayward metric, together with the resulting corrections to $\tau_*/L_*$, is provided in \ref{app:weyl}.

A second robustness issue concerns dynamical perturbations and anisotropies in realistic collapse or rotating configurations. Generic perturbations lift the degeneracy of the tidal spectrum, producing distinct eigenvalues $\lambda_1\neq\lambda_2\neq\lambda_3$. The operational timescale remains $\tau_*=|\lambda_{\max}|^{-1/2}$ but is now set by the fastest tidal channel. This strengthens the operational interpretation: $\tau_*$ identifies the earliest proper time on which any curvature-induced geometric evolution becomes resolvable.

A third point, central for regular black holes, is the well-known instability of inner Cauchy horizons. Perturbations propagating into the interior generically undergo exponential blueshift, triggering the mass-inflation instability and potentially altering the internal geometry~\cite{PoissonIsrael1990,CarballoRubio2018}. The local operational statements of sections~\ref{sec:operational_scales} through~\ref{sec:perturbation_transfer} depend only on the local tidal tensor along a freely falling worldline and are therefore unaffected. Global inferences that rely on quasi-stationary interior properties, in particular calibration strategies that fix the core scale $\ell$ from an extremality condition, should be understood as conditional on effective regulation of mass inflation.

\paragraph{Connection to energy condition violation.}
\label{subsec:nec_epsC}
The anisotropy parameter $\epsilon_C$ acquires additional physical meaning through its connection to the null energy condition (NEC). For the Hayward metric, the effective stress-energy tensor satisfies $\rho+p_\perp=(8\pi)^{-1}(\mathcal{A}+4\mathcal{B}-\mathcal{F})$, where $\mathcal{A}$, $\mathcal{B}$, $\mathcal{F}$ are the orthonormal-frame curvature components defined in \ref{app:weyl}. In the de~Sitter core one has $\rho+p_\perp=-6/(8\pi\ell^2)<0$ (maximal NEC violation), whereas at the horizon $\rho+p_\perp\to 0$. The NEC violation extends from $r=0$ to the horizon and decreases monotonically outward. The radial NEC ($\rho+p_r\propto R_{00}+R_{11}=0$) is identically saturated for any static spherically symmetric metric; therefore the angular NEC is the only non-trivial condition. Both $\epsilon_C$ and the NEC violation are controlled by the departure from de~Sitter geometry and both vanish at the same boundaries, but they measure distinct aspects: $\epsilon_C$ quantifies the tidal anisotropy, while $\rho+p_\perp$ measures the total effective stress-energy departure from the de~Sitter equation of state. Their shared boundary structure reflects the common geometric origin: the NEC-violating effective matter that sustains the regular core necessarily introduces Weyl anisotropy in the transition region.

\section{Discussion, conclusions, and outlook}
\label{sec:discussion}

\subsection{Physical interpretation}
\label{subsec:interpretation}

The operational scales derived in this study do not imply that spacetime becomes discrete below $L_*$ or that proper time acquires a minimal quantum. The underlying manifold remains smooth and Lorentz invariance is preserved. What changes in the high-curvature regime is the accuracy-dependent domain of validity of the local-inertial approximation and the proper-time resolution of tidal dynamics.

It is useful to compare this geometric bound to the familiar measurement-backreaction intuition. A standard dimensional estimate for localizing a probe to a length $\Delta x$ requires a characteristic energy $E\sim\hbar c/\Delta x$, yielding an energy density $\rho_{\rm probe}\sim\hbar c/(\Delta x)^4$. This estimate becomes comparable to the curvature-limited density $\rho\sim K_{\max}\sim L_*^{-4}$ precisely when $\Delta x$ approaches $L_*$. In frameworks where curvature is dynamically bounded,
localization attempts beyond $L_*$ probe a regime in which
the tidal response is necessarily nonperturbative.

More broadly, these results suggest a reinterpretation of the Planck scale. Rather than a boundary beyond which classical spacetime loses meaning, the Planckian regime can be viewed as a transition to a high-curvature phase in which tidal dynamics is maximal but finite and the local-inertial approximation becomes accuracy limited. The geometry remains regular and causal structure is preserved. This does not remove the need for quantum gravity, but it sharpens its scope: once curvature and tides are classically regulated, the remaining tasks for a quantum theory genuinely concern quantum aspects such as microstates, entanglement, and unitarity, rather than singularity removal.

\subsection{Limitations}
\label{subsec:limitations}

The conclusions of this study are local and kinematic: they follow from bounds on tidal observables and do not constitute a microscopic completion of gravity.

The benchmark coefficient $\tau_*/L_*=24^{1/4}$ is exact only for near-conformally flat, maximally symmetric cores. In generic interiors with non-negligible Weyl curvature the coefficient receives corrections controlled by $\epsilon_C$ as quantified in section~\ref{sec:robustness} and \ref{app:weyl}.

Regular black hole geometries with de~Sitter-like cores generically feature an inner Cauchy horizon. Classical and semiclassical analyses indicate that perturbations undergo strong blueshift at such horizons, leading to mass inflation and potentially substantial modification of the interior structure~\cite{PoissonIsrael1990,CarballoRubio2018}. This motivates the treatment extremality-based calibrations as
conditional, as discussed in section~\ref{sec:robustness}.

The critical scale $k_*\sim\tau_*^{-1}$ identifies which perturbation modes are sensitive to the high-curvature phase, but quantitative predictions for the scalar spectral index $n_s$, the tensor-to-scalar ratio $r$, and non-Gaussianity require solving the Mukhanov-Sasaki equation on a specific background and specifying a quantum state.

Finally, replacing a curvature singularity by a regular core does not by itself resolve the black hole information problem, which involves quantum dynamics of entanglement and unitarity associated with horizons and evaporation. The renormalized trace anomaly
$\langle T^\mu{}_\mu\rangle_{\rm ren}$ is quadratic in curvature invariants, with contributions from the squared Weyl tensor $C^2$, the Gauss--Bonnet density $E_4$, and $\Box R$~\cite{Birrell_Davies}. Under the bounded-curvature hypothesis each of these terms satisfies $C^2,\,E_4,\,(\Box R)\lesssim\mathcal{O}(\ell^{-4})$, so the renormalized stress tensor remains finite throughout the interior, in contrast to the Schwarzschild case where $\langle T_{\mu\nu}\rangle_{\rm ren}$ diverges as $r\to 0$.

\subsection{Summary of results}
\label{subsec:summary}

We studied dynamical consequences of bounded curvature in a covariant, classical setting. Given an invariant bound on the tidal eigenvalues along freely falling worldlines, we established two model-independent results. The deformation bound of section~\ref{subsec:deformation_bound} provides a rigorous upper bound on accumulated geodesic deviation through any bounded curvature interior, with $\tau_*$ as the characteristic timescale. Explicit integration through the extremal Hayward geometry confirmed this bound and demonstrated geodesic completeness of the regular interior. The adiabaticity bound of section~\ref{subsec:adiabaticity} establishes a critical wavenumber $k_*\sim\tau_*^{-1}$ separating adiabatic from non-adiabatic perturbation transfer, with Bogoliubov coefficients exponentially suppressed for $k\,\tau_*\gg 1$. Both results depend only on the tidal bound; the first controls classical deformation of geodesic bundles, the second controls quantum-mechanical mode mixing of field perturbations.

The results apply to any theory with bounded curvature, including action-level limiting-curvature theories~\cite{Frolov2021_Bounce,Frolov2022_4D} and density-responsive gravity~\cite{Drobczyk2025DRG}; quantitative calibration within specific models is left to future work.

\subsection{Outlook}
\label{subsec:outlook}

Several directions follow naturally from this study.

\emph{First}, the critical non adiabaticity scale $k_*\sim\tau_*^{-1}$ should be quantified for specific background profiles by solving the Mukhanov-Sasaki equation on explicit bounce or core-transit backgrounds. This would determine how the perturbation spectra ($n_s$, $r$, $f_{\rm NL}$) depend on $k/k_*$ and identify which observables are sensitive to the bounded curvature phase.

\emph{Second}, the calibration of $\ell$ should be stress-tested against realistic interior dynamics. Regular black holes with inner horizons can exhibit mass-inflation instabilities~\cite{PoissonIsrael1990,CarballoRubio2018}, and a focused analysis should determine whether bounded tidal eigenvalues persist in dynamical collapse solutions and under perturbations.

\emph{Third}, the operational scales connect directly to the gravitational wave phenomenology. Regular core geometries introduce an interior cavity whose round-trip time $\Delta t_{\rm echo}\sim 2\!\int_{r_-}^{r_+}\!dr\,|f|^{-1}$ provides a geometry-dependent prediction for echo delays~\cite{Cardoso2016_Echoes}. Bounded curvature cores also generically induce nonzero tidal Love numbers, in contrast to the vanishing Love numbers of classical four-dimensional black holes~\cite{Binnington_Poisson_2009}, offering a potential gravitational-wave observable sensitive to the core structure at the scale $\ell$.

\emph{Fourth}, a careful comparison of the tidal-limited geometric resolvability discussed here with quantum-limited measurement analyses for rods and clocks~\cite{Salecker1958,Brukner2015} would clarify where classical curvature controlled limitations on local inertial frames coincide with quantum uncertainties of measurement devices and where they differ.

\emph{Fifth}, the perturbation transfer scale $k_*\sim\tau_*^{-1}$ derived in section~\ref{sec:perturbation_transfer} governs which modes retain coherence across a cosmological bounce and which undergo non-adiabatic mixing. In frameworks where bounded curvature guarantees a non-singular bounce, this provides a concrete starting point for analysing perturbation spectra in cyclic cosmologies.

\section*{Data availability}
The numerical code and data that support the findings of this article are openly available at
Ref.~\cite{Drobczyk2026_code}.

\appendix

\section{Curvature Invariants for the Hayward Metric}
\label{app:curvature}

In this appendix, we derive the curvature invariants for the Hayward metric and rigorously establish that the Kretschmann scalar attains its maximum at $r=0$ and decreases monotonically outward.

In this appendix, we use the curvature convention
$R^\rho{}_{\sigma\mu\nu}=\partial_\mu\Gamma^\rho_{\nu\sigma}-\partial_\nu\Gamma^\rho_{\mu\sigma}
+\Gamma^\rho_{\mu\lambda}\Gamma^\lambda_{\nu\sigma}-\Gamma^\rho_{\nu\lambda}\Gamma^\lambda_{\mu\sigma}$,
metric signature $(-,+,+,+)$, and an orthonormal tetrad
$e^{\hat t}=\sqrt{f}\,dt$, $e^{\hat r}=f^{-1/2}\,dr$, $e^{\hat\theta}=r\,d\theta$,
$e^{\hat\phi}=r\sin\theta\,d\phi$.
With these choices the component expressions below follow straightforwardly from the Cartan structure equations.

\subsection{Setup and Metric Components}
\label{app:setup}

The Hayward metric is given by equation~(\ref{eq:Hayward}) with the metric function
\begin{equation}
    f(r) = 1 - \frac{2Mr^2}{r^3 + 2M\ell^2}.
    \label{eq:f_app}
\end{equation}
For computational clarity, we introduce the dimensionless variables
\begin{equation}
    x \equiv \frac{r}{\ell}, \qquad m \equiv \frac{M}{\ell},
    \label{eq:dimensionless}
\end{equation}
so that
\begin{equation}
    f(x) = 1 - \frac{2m x^2}{x^3 + 2m}.
    \label{eq:f_dimensionless}
\end{equation}
The non-vanishing metric components in coordinates $(t, r, \theta, \phi)$ are
\begin{equation}
    g_{tt} = -f(r), \quad g_{rr} = f(r)^{-1}, \quad g_{\theta\theta} = r^2, \quad g_{\phi\phi} = r^2\sin^2\theta.
    \label{eq:metric_components}
\end{equation}

\subsection{Riemann Tensor Components}
\label{app:riemann}

For a static, spherically symmetric metric, the independent non-vanishing components of the Riemann tensor (in an orthonormal frame) are:
\begin{align}
    R^{\hat{t}}_{\ \hat{r}\hat{t}\hat{r}} &= -\frac{1}{2}f''(r), \label{eq:R_trtr} \\
    R^{\hat{t}}_{\ \hat{\theta}\hat{t}\hat{\theta}} &= R^{\hat{t}}_{\ \hat{\phi}\hat{t}\hat{\phi}} = -\frac{f'(r)}{2r}, \label{eq:R_tthetattheta} \\
    R^{\hat{r}}_{\ \hat{\theta}\hat{r}\hat{\theta}} &= R^{\hat{r}}_{\ \hat{\phi}\hat{r}\hat{\phi}} = -\frac{f'(r)}{2r}, \label{eq:R_rtheta} \\
    R^{\hat{\theta}}_{\ \hat{\phi}\hat{\theta}\hat{\phi}} &= \frac{1 - f(r)}{r^2}. \label{eq:R_thetaphi}
\end{align}

Here, primes denote derivatives with respect to $r$, and hatted indices refer to the orthonormal frame $e^{\hat{t}} = \sqrt{f}\,dt$, $e^{\hat{r}} = f^{-1/2}dr$, etc.

Note that the components listed here use mixed-index notation ($R^{\hat a}{}_{\ \hat b\hat c\hat d}$); the all-lower-index convention adopted in~\ref{app:weyl} differs by a sign for components involving $\hat{0}$ (cf.\ the footnote in~\ref{app:weyl_C2}).

\subsection{Derivatives of the Metric Function}
\label{app:derivatives}

From equation~(\ref{eq:f_app}), direct differentiation yields the first derivative:
\begin{equation}
    f'(r) = \frac{2Mr(r^3 - 4M\ell^2)}{(r^3 + 2M\ell^2)^2}.
    \label{eq:f_prime}
\end{equation}
Differentiating again and simplifying gives the second derivative:
\begin{equation}
    f''(r) = \frac{-4M(r^6 - 14M\ell^2 r^3 + 4M^2\ell^4)}{(r^3 + 2M\ell^2)^3}.
    \label{eq:f_double_prime}
\end{equation}

\subsection{Behavior at the Origin}
\label{app:origin}

Taking limits as $r \to 0$:
\begin{equation}
    f(0) = 1, \quad f'(0) = 0, \quad f''(0) = -\frac{2}{\ell^2}.
\end{equation}
Thus, near $r = 0$:
\begin{equation}
    f(r) \approx 1 - \frac{r^2}{\ell^2} + \mathcal{O}(r^4),
    \label{eq:f_near_origin}
\end{equation}
confirming the de~Sitter form with effective cosmological constant $\Lambda_{\mathrm{eff}} = 3/\ell^2$.

\subsection{Kretschmann Scalar and Monotonicity}
\label{app:kretschmann}

The Kretschmann scalar is defined as $K = R_{\mu\nu\rho\sigma}R^{\mu\nu\rho\sigma}$. For our metric:
\begin{equation}
  K = \left(f''\right)^2 + \frac{4(f')^2}{r^2} + \frac{4(1-f)^2}{r^4}.
    \label{eq:K_formula}
\end{equation}
Using equations~(\ref{eq:f_prime}) and (\ref{eq:f_double_prime}) in equation~(\ref{eq:K_formula}), the Kretschmann scalar can be written in closed form as:
\begin{equation}
    K(r)=
    \frac{48M^2\left(r^{12}-8M\ell^2 r^9+72M^2\ell^4 r^6-16M^3\ell^6 r^3+32M^4\ell^8\right)}
    {\left(r^3+2M\ell^2\right)^6}.
    \label{eq:K_closed}
\end{equation}
As cross-checks, equation~(\ref{eq:K_closed}) reproduces (i) the de~Sitter value $K(0)=24/\ell^4$ and (ii) the Schwarzschild falloff $K(r)\to 48M^2/r^6$ for $r\gg (M\ell^2)^{1/3}$.

At the origin ($r=0$), this evaluates to:
\begin{equation}
  K(0) = \frac{48M^2(32M^4\ell^8)}{(2M\ell^2)^6} = \frac{1536M^6\ell^8}{64M^6\ell^{12}} = \frac{24}{\ell^4} = \Kmax.
    \label{eq:K_max_result}
\end{equation}

We now rigorously prove that $K(r)$ decreases is strictly monotonically for $r>0$. Differentiating equation~(\ref{eq:K_closed}) and factorizing, one obtains:
\begin{equation}
 \frac{dK}{dr}=
    -\frac{288\,M^2\,r^2}{\left(r^3+2M\ell^2\right)^7}\,
    \Big(r^{12}-16M\ell^2 r^9+168M^2\ell^4 r^6-184M^3\ell^6 r^3+112M^4\ell^8\Big).
    \label{eq:dK_dr_correct}
\end{equation}
For $r>0$, the prefactor is strictly negative. It remains to show that the polynomial in parentheses is strictly positive. Introduce the dimensionless variable $y \equiv r^3/(M\ell^2) > 0$. Then the term in parentheses becomes $M^4\ell^8\,q(y)$, with
\begin{equation}
    q(y)=y^4-16y^3+168y^2-184y+112.
    \label{eq:q_polynomial}
\end{equation}

We note that $q(0)=112>0$. A more robust argument follows from the stationary-point structure of $q(y)$.
Differentiating,
\begin{equation}
q'(y)=4y^3-48y^2+336y-184 = 4\big(y^3-12y^2+84y-46\big).
\end{equation}
The discriminant of the cubic in parentheses is negative ($\Delta=-895212<0$), hence it has exactly one real root $y_0$.
Therefore $q(y)$ possesses a single global minimum at $y_0$, and it is sufficent to evaluate $q(y_0)$.
Numerically one finds $y_0\simeq 0.596$ and $q(y_0)\simeq 58.8>0$, implying $q(y)>0$ for all $y\ge 0$.
Consequently,
\begin{equation}
\frac{dK}{dr}<0\qquad \forall\, r>0,
\end{equation}
and $K(r)$ decreases strictly monotonically from its global maximum $K(0)=24/\ell^4$ toward $K(r)\to 0$ as $r\to\infty$.

\section{Weyl Decomposition and Robustness for the Hayward Metric}
\label{app:weyl}

In this appendix we quantify the departure from conformal flatness in the Hayward geometry, providing an explicit realization of the robustness estimates discussed in section~\ref{sec:robustness}. Throughout, we use orthonormal-frame components with \emph{all lower indices} and the curvature-sign conventions specified in~\ref{app:curvature}.

\subsection{Weyl-squared scalar for static spherically symmetric metrics}
\label{app:weyl_C2}

For a static, spherically symmetric line element with lapse function $f(r)$ (equation~(\ref{eq:f_app})), the six independent orthonormal-frame Riemann components reduce to three independent functions,\footnote{%
These are all-lower-index orthonormal-frame components. The sign of $R_{\hat 0\hat 1\hat 0\hat 1}$ differs from the mixed-index expression $R^{\hat 0}{}_{\hat 1\hat 0\hat 1}=-f''/2$ commonly quoted in the literature by a factor $g_{\hat 0\hat 0}=-1$.}
\begin{align}
\mathcal{A} &\equiv R_{\hat 0\hat 1\hat 0\hat 1} = \frac{1}{2}f'',  \label{eq:calA}\\
\mathcal{B} &\equiv R_{\hat 0\hat 2\hat 0\hat 2} = R_{\hat 0\hat 3\hat 0\hat 3} = \frac{f'}{2r}, \label{eq:calB}\\
\mathcal{F} &\equiv R_{\hat 2\hat 3\hat 2\hat 3} = \frac{1-f}{r^2}, \label{eq:calF}
\end{align}
with $R_{\hat 1\hat 2\hat 1\hat 2}=R_{\hat 1\hat 3\hat 1\hat 3}=-\mathcal{B}$.
The tidal eigenvalues of $E_{ij}=R_{\hat 0 i \hat 0 j}$ are $\lambda_1=\mathcal{A}$ (radial) and $\lambda_2=\lambda_3=\mathcal{B}$ (angular), and the Kretschmann scalar is
\begin{equation}
K = 4\mathcal{A}^2 + 16\mathcal{B}^2 + 4\mathcal{F}^2.
\label{eq:K_ABC}
\end{equation}

Using the Gauss decomposition
$C_{\mu\nu\rho\sigma}C^{\mu\nu\rho\sigma}=K-2R_{\mu\nu}R^{\mu\nu}+\frac{1}{3}R^2$
together with the orthonormal-frame Ricci components
\begin{equation}
R_{\hat 0\hat 0}=\mathcal{A}+2\mathcal{B},\quad
R_{\hat 1\hat 1}=-\mathcal{A}-2\mathcal{B},\quad
R_{\hat 2\hat 2}=R_{\hat 3\hat 3}=-2\mathcal{B}+\mathcal{F},
\end{equation}
one finds after a short algebra
\begin{equation}
C_{\mu\nu\rho\sigma}C^{\mu\nu\rho\sigma}
= \frac{4}{3}\!\left(\mathcal{A} - 2\mathcal{B} - \mathcal{F}\right)^2.\ 
\label{eq:C2_general_SSS}
\end{equation}
Hence $C^2\equiv C_{\mu\nu\rho\sigma}C^{\mu\nu\rho\sigma}$ vanishes if and only if
$\mathcal{A}=2\mathcal{B}+\mathcal{F}$, which is precisely the maximally symmetric (conformally flat) condition for a static spherically symmetric geometry.

As a consistency check, in vacuum ($R_{\mu\nu}=0$, implying $\mathcal{A}+2\mathcal{B}=0$ and $\mathcal{F}=2\mathcal{B}$) one has
$\mathcal{A}-2\mathcal{B}-\mathcal{F}=-6\mathcal{B}$ and thus $C^2=48\mathcal{B}^2$; for Schwarzschild ($\mathcal{B}=-M/r^3$) this reproduces $C^2=48M^2/r^6=K$ as required.

\subsection{Closed-form expressions for the Hayward metric}
\label{app:weyl_Hayward}

For the Hayward lapse, using $D\equiv r^3+2M\ell^2$ for brevity, one obtains
\begin{equation}
\mathcal{A}-2\mathcal{B}-\mathcal{F}
= -\frac{6Mr^3(r^3 - 4M\ell^2)}{D^3}\,,
\label{eq:combo_Hayward}
\end{equation}
and therefore
\begin{equation}
\ C^2(r)
= \frac{48\,M^2 r^6\,(r^3 - 4M\ell^2)^2}{(r^3 + 2M\ell^2)^6}\,.\ 
\label{eq:C2_Hayward}
\end{equation}

\paragraph{Asymptotic checks.}
\begin{itemize}
\item \emph{Core ($r\to 0$):} The prefactor $r^6$ ensures $C^2(0)=0$, confirming that the de~Sitter core is conformally flat.
\item \emph{Schwarzschild limit ($r\gg(M\ell^2)^{1/3}$):} One finds $C^2\to 48M^2/r^6$, matching the vacuum Kretschmann scalar, that is \ $C^2/K\to 1$.
\item \emph{Intermediate radius ($r^3=4M\ell^2$, i.e.\ $y=4$):} $C^2=0$, i.e.\ the geometry becomes momentarily conformally flat at this radius (equivalently, where $f'(r)=0$).
\end{itemize}

\subsection{The anisotropy parameter $\epsilon_C(r)$}
\label{app:weyl_epsC}

The dimensionless Weyl-to-Kretschmann ratio introduced in equation~(\ref{eq:epsC_def}),
\begin{equation}
\epsilon_C^2(r)\equiv \frac{C^2(r)}{K(r)}\,,\qquad 0\le \epsilon_C \le 1,
\end{equation}
is most conveniently expressed in terms of the dimensionless radial variable $y\equiv r^3/(M\ell^2)$:
\begin{equation}
\epsilon_C^2(y)
= \frac{y^2\,(y-4)^2}{y^4 - 8y^3 + 72y^2 - 16y + 32}\,.
\label{eq:epsC_Hayward}
\end{equation}
This function satisfies $\epsilon_C(0)=0$ (maximally symmetric core) and $\epsilon_C(4)=0$ (the conformally flat point where $f'=0$), while $\epsilon_C\to 1$ as $y\to\infty$ where the geometry approaches Schwarzschild vacuum and $K=C^2$. Note that in this far-exterior regime the curvature invariants themselves vanish, so the operational bounds of the main text become physically irrelevant despite $\epsilon_C\to 1$.

Within the deep interior $0<y<4$, $\epsilon_C$ has a single maximum at $y=1$, where $\epsilon_C(1)=1/3$. For $y>4$, $\epsilon_C$ increases monotonically from $0$ toward $1$.

\subsection{Tidal eigenvalues and operational ratio away from the core}
\label{app:weyl_tidal}

For the Hayward metric, the two distinct tidal eigenvalues (in units of $\ell^{-2}$) are
\begin{equation}
\lambda_1\,\ell^2
= \frac{-2(y^2 - 14y + 4)}{(y+2)^3}\,,
\qquad
\lambda_2\,\ell^2
= \frac{y - 4}{(y+2)^2}\,,
\label{eq:tidal_eigenvalues_y}
\end{equation}
with $\lambda_{\max}\equiv\max(|\lambda_1|,|\lambda_2|)$. Both eigenvalues are degenerate at $y=0$,
$\lambda_1=\lambda_2=-\ell^{-2}$, consistent with the de~Sitter relation $R_{\hat 0 i\hat 0 j}=-\ell^{-2}\delta_{ij}$ (defocusing). The radial eigenvalue $\lambda_1$ changes sign at $y=7-3\sqrt{5}\approx 0.29$, marking the transition from de~Sitter-like defocusing to Schwarzschild-like focusing in the radial direction; the angular eigenvalue $\lambda_2$ changes sign later at $y=4$.

Defining the local operational ratio by
\begin{equation}
\frac{\taustar}{\Lstar}(r)\ \equiv\ \frac{K(r)^{1/4}}{|\lambda_{\max}(r)|^{1/2}}\,,
\end{equation}
one may evaluate $\taustar/\Lstar$ as a function of $y$; Table~\ref{tab:robustness} lists representative values.

\begin{table}[t]
\caption{Anisotropy parameter $\epsilon_C$, tidal eigenvalues, and operational ratio $\taustar/\Lstar$
as functions of $y=r^3/(M\ell^2)$ for the Hayward metric.
The maximally symmetric benchmark $\taustar/\Lstar=24^{1/4}\approx 2.213$ is recovered at $y=0$.}
\label{tab:robustness}
\centering
\begin{tabular}{cccccc}
\hline
$y$ & $\epsilon_C$ & $\lambda_1\,\ell^2$ & $\lambda_2\,\ell^2$ & $\taustar/\Lstar$ & Deviation \\
\hline
 0      & 0     & $-1.000$ & $-1.000$ & 2.213 & $0\%$ \\
 0.01   & 0.007 & $-0.951$ & $-0.988$ & 2.208 & $-0.2\%$ \\
 0.1    & 0.070 & $-0.564$ & $-0.884$ & 2.172 & $-1.9\%$ \\
 0.5    & 0.273 & $+0.352$ & $-0.560$ & 2.253 & $+1.8\%$ \\
 1      & 0.333 & $+0.667$ & $-0.333$ & 1.861 & $-15.9\%$ \\
 2      & 0.258 & $+0.625$ & $-0.125$ & 1.638 & $-26.0\%$ \\
 $y_\star\!\approx\!2.12$ & 0.245 & $+0.606$ & $-0.111$ & 1.633 & $-26.2\%$ \\
 4      & 0     & $+0.333$ & $0$      & 1.682 & $-24.0\%$ \\
 8      & 0.476 & $+0.088$ & $+0.040$ & 2.300 & $+3.9\%$ \\
 50     & 0.987 & $-0.026$ & $+0.017$ & 2.115 & $-4.4\%$ \\
 200    & 0.999 & $-0.009$ & $+0.005$ & 1.911 & $-13.7\%$ \\
\hline
\end{tabular}
\end{table}

Several features deserve comment:
\begin{enumerate}
\item \emph{Deep core ($y\lesssim 0.1$):} The benchmark $24^{1/4}$ is reproduced to better than $2\%$, confirming the stability of the maximally symmetric value in the region where bounded-curvature/operational arguments are intended to apply.
\item \emph{Intermediate interior ($y\sim 1{-}4$):} The ratio can deviate by up to ${\sim}26\%$ near $y_\star\approx 2.12$ (the interior point of maximal $\tau_*/L_*$ deviation from
the benchmark; the degenerate horizon of the extremal
Hayward configuration lies at $y=4$). This explicitly realizes the $\mathcal{O}(\epsilon_C)$ corrections discussed in section~\ref{sec:robustness} and shows that $24^{1/4}$ is a benchmark coefficient rather than a universal constant across the entire interior. The sign change of $\lambda_1$ at $y\approx 0.29$ reflects the transition from the de~Sitter-like defocusing core to the focusing region that connects to the Schwarzschild exterior.
\item \emph{Exterior ($y\gg 4$):} The geometry approaches Schwarzschild vacuum, $\epsilon_C\to 1$ and $K=C^2$. However, because $K\to 0$ in the far exterior, the operational bounds become physically irrelevant there.
\end{enumerate}
\begin{figure}[t]
\centering
\includegraphics[width=0.7\linewidth]{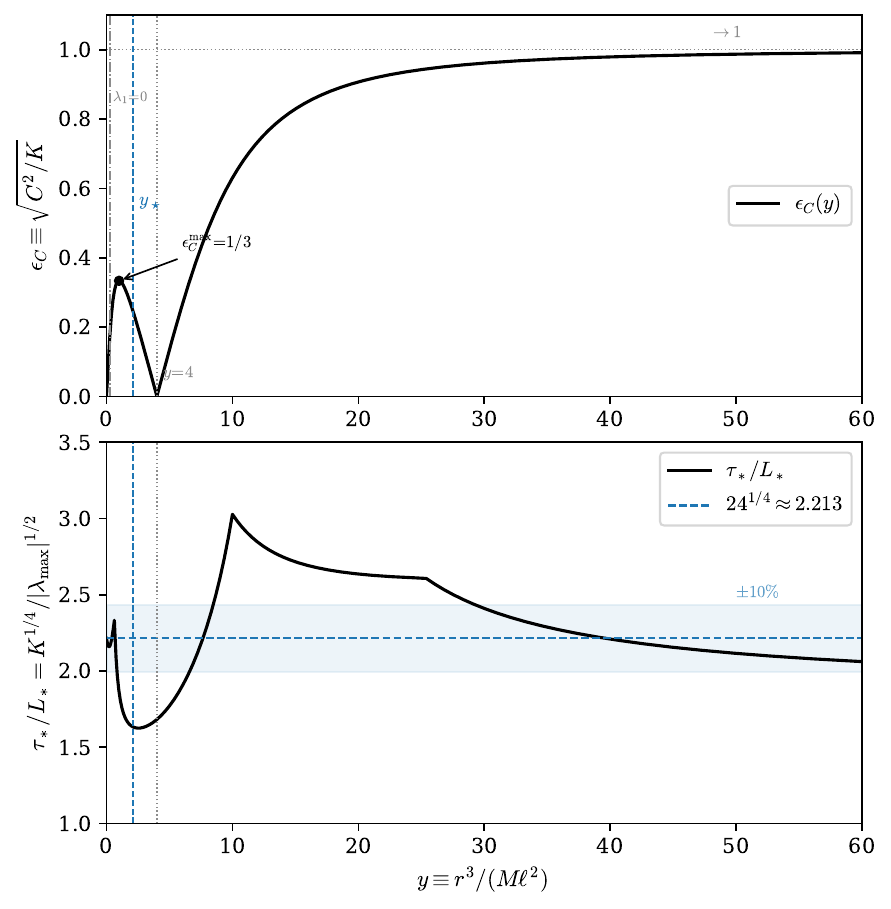}
\caption{Radial profiles of the Weyl anisotropy parameter $\epsilon_C(y)$ (top) and the operational ratio $\taustar/\Lstar(y)$ (bottom) for the Hayward metric, with $y\equiv r^3/(M\ell^2)$. In the top panel, $\epsilon_C=0$ at the de~Sitter core ($y=0$) and at $y=4$ (where $f'(r)=0$), and $\epsilon_C\to 1$ in the Schwarzschild limit ($y\to\infty$) where $K=C^2$. The gray dash-dotted line marks the sign change of $\lambda_1$ at $y\approx 0.29$. The dashed horizontal line in the lower panel marks the maximally symmetric benchmark $24^{1/4}$, and the shaded band indicates ${\pm}10\%$. The blue dashed vertical line marks $y\approx 2.12$, the point of maximal $\tau_*/L_*$ deviation; the extremal horizon lies at $y=4$ (where $\epsilon_C=0$).}
\label{fig:robustness_epsC}
\end{figure}
Figure~\ref{fig:robustness_epsC} displays the radial profiles of $\epsilon_C(y)$ and $\taustar/\Lstar(y)$.

\bibliography{main}

\end{document}